\documentclass[12pt]{article}
\usepackage[latin1]{inputenc}
\usepackage[dvips]{graphicx,color}
\usepackage{amssymb}
\usepackage{bibspacing}

\def\beq{\begin{equation}}
\def\eeq{\end{equation}}
\def\eq{\end{equation}}
\def\ba{\begin{eqnarray}}
\def\ea{\end{eqnarray}}
\def\Dslash{\slash \!\!\!\!D}

\def\centeron#1#2{{\setbox0=\hbox{#1}\setbox1=\hbox{#2}\ifdim
\wd1>\wd0\kern.5\wd1\kern-.5\wd0\fi \copy0\kern-.5\wd0\kern-.5\wd1\copy1\ifdim\wd0>\wd1 \kern.5\wd0\kern-.5\wd1\fi}}
\def\ltap{\;\centeron{\raise.35ex\hbox{$<$}}{\lower.65ex\hbox{$\sim$}}\;}
\def\gtap{\;\centeron{\raise.35ex\hbox{$>$}}{\lower.65ex\hbox{$\sim$}}\;}

\newcommand{\captionfonts}{\small}
\makeatletter \long\def\@makecaption#1#2{
  \vskip\abovecaptionskip
  \sbox\@tempboxa{{\captionfonts #1: #2}}
  \ifdim \wd\@tempboxa >\hsize
    {\captionfonts #1: #2\par}
  \else
    \hbox to\hsize{\hfil\box\@tempboxa\hfil}
    \fi
  \vskip\belowcaptionskip}
\makeatother

\newcommand{\newc}{\newcommand}
\newc{\qbar}{{\overline q}}
\newc{\Kahler}{K\"ahler }
\newc{\deltaGS}{\delta_{\rm GS}}
\newc{\bsg}{B\rightarrow X_s\gamma}
\newc{\Bmumu}{B_s\rightarrow \mu^+ \mu^-}
\newc{\MSusy}{m_{\tilde{t}}}
\newc{\HEWSB}{H^{\textrm{\tiny{EWSB}}}}
\newc{\cossqbma}{\cos^2(\beta-\alpha)}
\newc{\sinsqbma}{\sin^2(\beta-\alpha)}
\newc{\HSM}{H_{\textrm{\tiny{SM}}}}
\newc{\MSusyMin}{m_{\tilde{t},\textrm{\tiny{min}}}}
\newc{\MGUT}{M_{\mbox{\scriptsize{GUT}}}}

\begin{document}
\def\pplogo{\vbox{\kern-\headheight\kern -29pt
\halign{##&##\hfil\cr&{\ppnumber}\cr\rule{0pt}{2.5ex}&\ppdate\cr}}} \makeatletter
\def\ps@firstpage{\ps@empty \def\@oddhead{\hss\pplogo}
  \let\@evenhead\@oddhead
}
\def\maketitle{\par
 \begingroup
 \def\thefootnote{\fnsymbol{footnote}}
 \def\@makefnmark{\hbox{$^{\@thefnmark}$\hss}}
 \if@twocolumn
 \twocolumn[\@maketitle]
 \else \newpage
 \global\@topnum\z@ \@maketitle \fi\thispagestyle{firstpage}\@thanks
 \endgroup
 \setcounter{footnote}{0}
 \let\maketitle\relax
 \let\@maketitle\relax
 \gdef\@thanks{}\gdef\@author{}\gdef\@title{}\let\thanks\relax}
\makeatother

%%%%%%%%%%%%%%%%%%%%%%%%%%%%%%%%%%%%%%%%%%%%%%%%%%%%%%%%%%%%%%

\setcounter{page}0
\def\ppnumber{\vbox{\baselineskip14pt
%\hbox{hep-ph/0000000}
}}
\def\ppdate{RUNHETC-2007-20} \date{}

\author{\\Rouven Essig\footnote{rouven@physics.rutgers.edu} \\
[7mm]
{\normalsize NHETC, Department of Physics and Astronomy,}\\
{\normalsize Rutgers University, Piscataway, NJ 08854, U.S.A.}\\}

\title{\bf \LARGE Direct Detection of Non-Chiral \\Dark Matter}
\maketitle
\vskip 2cm

\begin{abstract} \normalsize
\noindent
Direct detection experiments rule out fermion dark matter that is a chiral
representation of the electroweak gauge group.
Non-chiral real, complex and singlet representations, however, provide
viable fermion dark matter candidates.
Although any one of these candidates will be virtually impossible to
detect at the LHC, it is shown that they may be detected at future planned
direct detection experiments.
For the real case, an irreducible radiative coupling to quarks may allow
a detection.
The complex case in general has an experimentally ruled out tree-level coupling
to quarks via $Z$-boson exchange.
However, in the case of two $SU(2)_L$ doublets, a higher dimensional coupling
to the Higgs can suppress this coupling, and a remaining irreducible radiative
coupling may allow a detection.
Singlet dark matter could be detected through a coupling to quarks via Higgs
exchange.
Since all non-chiral dark matter can have a coupling to the Higgs, at least some
of its mass can be obtained from electroweak symmetry breaking, and this mass is
a useful characterization of its direct detection cross-section.
\end{abstract}
\bigskip

\newpage

\tableofcontents

\section{Introduction}\label{Sec:Introduction}

The evidence for the existence of non-baryonic dark matter is
overwhelming.
Within the concordance $\Lambda$CDM cosmological model, the
required dark matter relic density is now known to remarkable
accuracy \cite{Spergel:2006hy}.
The nature of the dark matter particles within this model,
however, is unknown.

There is a possibility that new physics associated with
electroweak symmetry breaking (EWSB) might contain a dark matter
candidate with the correct relic density.
This is because \emph{weakly interacting massive particles} (WIMPs)
can have the observed dark matter relic density through thermal
freeze-out if their mass is on the order of the electroweak (EW) scale.
In addition, it is possible to stabilize WIMPs by including a symmetry
that forbids their decay into other particles.
This allows them to be good dark matter candidates.

The preferred mass of WIMPs suggests the possibility that they
may be produced and detected at the upcoming Large Hadron Collider
(LHC) at CERN.
Two other types of experiments attempting to detect dark matter are indirect and
direct detection experiments.
While the indirect detection experiments look for the particles that are produced
from annihilating dark matter, the direct detection experiments attempt to infer the
presence of dark matter particles as they scatter off nuclei within detectors by
looking for the resulting nuclear recoil.

The rationale for the direct detection experiments is that the dark matter lies in a
halo which encompasses our Milky Way galaxy.
As the earth and sun rotate around the galactic center, detectors on
the earth move through the halo and intersect the path of dark
matter particles, which are expected to scatter off the nuclei inside
the detectors.
Since the local dark matter density is not known better than to within a
factor of two, there is some uncertainty in the expected scattering rate
\cite{Jungman:1995df}.
Depending on the experimental setup, the nuclear recoil from the
scattering would produce ionization, phonons or scintillation, any
of which can be observed.
Examples of direct dark matter detection experiments include CDMS, DAMA,
NaIAD, PICASSO, ZEPLIN, EDELWEISS, CRESST, XENON and WARP
\cite{Akerib:2005za,Akerib:2005kh,Bernabei:2000qi,Bernabei:2003za,
Ahmed:2003su,Barnabe-Heider:2003cq,Kudryavtsev:2004ju,Benoit:2002hf,
Angloher:2004tr,Aprile:2006kx,Benetti:2006az}.

The dark matter scattering off nuclei within a detector can proceed
via two fundamentally different types of interactions.
There is, on the one hand, a spin-independent,
or coherent, interaction between the dark matter and the nucleons.
In this case the contribution of each nucleon to the total
scattering cross-section interferes constructively across the
nucleus.
Scattering off nuclei is therefore enhanced roughly by a factor of
$A^2$ in the cross-section, where $A$ is the number of nucleons in
the nucleus.
This large enhancement factor is absent for the other type of
interaction, which is spin-dependent, and couples the dark matter
spin to the spin of the nuclei.
The large enhancement factor is also the main reason that much tighter
constraints (a factor of about $10^5-10^6$) exist on the SI
cross-section, normalized to cross-section per nucleon, than on the
SD cross-section.

In this paper, fermion dark matter transforming under the EW gauge group
$SU(2)_L$ $\times$ $U(1)_Y$ will be added to the standard model (SM), and
the observational consequences at a direct detection experiment will be discussed.
In particular, chiral and non-chiral (real and complex) representations of
$SU(2)_L$ $\times$ $U(1)_Y$ will be considered in \S \ref{Sec:Chiral DM} and
\S \ref{Sec:Non-chiral Dark Matter}, respectively, and the focus will be on
spin-independent interactions for the reasons discussed in the previous paragraph.
\S \ref{Sec:singlet DM} discusses how the direct detection cross-section
may be characterized in terms of the fraction of the dark matter mass that
is obtained through EWSB.
This characterization is particularly useful for EW singlet dark matter.
The conclusions are presented in \S \ref{Sec:Conclusion}.

The results of this paper are summarized in Figure \ref{Fig:Result}.
Shown are the current experimental upper bounds on the spin-independent
cross-section for WIMP scattering off nucleons from XENON10 (solid line)
\cite{Angle:2007uj}, the projected upper bounds for SuperCDMS 2-ST at
Soudan (blue dashed line), SuperCDMS 25kg / 7-ST at Snolab
(green dashed line), XENON1T (magenta dashed line) and SuperCDMS Phase C
(red dashed line) \cite{Aprile:2005mz,Brink:2005ej,Schnee:2005pj}.
The cross-sections for chiral and non-chiral dark matter are shown, in addition to
the Higgs contribution to the direct detection cross-section for a variety of
parameter choices.

\section{Chiral Electroweak Dark Matter}\label{Sec:Chiral DM}

Chiral EW matter is forbidden to have an explicit mass term in the
Lagrangian since such a mass term is not gauge invariant.
It instead has a Yukawa coupling to the Standard Model Higgs field
and gains all its mass from EWSB through the
Higgs mechanism.
Chiral EW dark matter particles are thus Dirac fermions.

EW precision measurements put tight constraints on
additional chiral matter.
For example, an additional doublet of colorless heavy fermions gives
a contribution of 1/$6\pi$ to the electroweak $S$-parameter, which is
about 1.8$\sigma$ away from its measured central value.
An additional degenerate generation is disfavored even more strongly at
the $99.95\%$ confidence level \cite{Eidelman:2004wy}.

Although EW precision measurements still allow room for
chiral EW dark matter, direct detection experiments rule it out as a
viable dark matter candidate.
The reason is that it has a vector coupling to the
$Z$-boson and can therefore scatter coherently off the nuclei inside
the detector via a tree-level $Z$-boson exchange.
The resulting cross-section is large enough that such dark matter
particles would already have been seen \cite{Goodman:1984dc}.

In general, the cross-section per nucleon for dark matter scattering
coherently off nuclei via the exchange of a $Z$-boson is given by

\beq \label{Eqn:Goodman-Witten} \sigma \simeq \frac{G_F^2}{2\pi}
m^2_{\chi N}\frac{1}{A^2}\Big[(1-4\sin^2\theta_W)Z - (A-Z)\Big]^2
\overline{Y}^2. \eeq

\noindent Here, $G_F$ is the Fermi coupling constant,
$m_{\chi N}$ is the reduced mass of the dark matter mass
($m_{\chi}$) and nucleon mass ($m_N$), $A$ ($Z$) is the mass (atomic) number
of the nucleus, $\theta_W$ is the weak mixing angle, and
$\overline{Y} \equiv \frac{1}{2}(Y_L + Y_R)$, where
$Y_L$ and $Y_R$ are the hypercharge of the left- and right-handed
components of the dark matter particle \cite{Goodman:1984dc}.
The convention chosen here is $Q = T_3 + \frac{1}{2} Y$, where $Q$ is
the electric charge, $T_3$ is the third component of the isospin, and
$Y$ is the hypercharge of the particle.
The term proportional to $Z$ in the square brackets is for the dark matter
scattering off the protons inside the nucleus.
It is suppressed since $1-4\sin^2\theta_W$ is very small.
The term proportional to $A-Z$ is for the dark matter scattering off the
neutrons inside the nucleus, and it dominates.
The factor of $1/A^2$ normalizes the cross-section to a cross-section
per nucleon.

Chiral EW dark matter has $Y_R = Y_L \pm 1$, i.e. $\overline{Y} = Y_L \pm \frac{1}{2}$.
For the CDMS experiment, for example, which uses Germanium ($^{73}_{32}$Ge), the
scattering cross-section per nucleon then becomes

\beq \sigma \gtrsim 5 \times 10^{-40} \textrm{ cm}^2, \eeq

\noindent for $\overline{Y}\ge\frac{1}{2}$.
This result is roughly independent of the mass of the dark matter, at least
for a large enough dark matter mass.
A Dirac neutrino\footnote{A Dirac neutrino also has an axial vector coupling
to the $Z$-boson and therefore a spin-dependent interaction with nuclei.}
saturates the lower bound as it has $Y_L = 1$ and $Y_R = 0$, and thus
$\bar{Y} = \frac{1}{2}$.
For $m_{\chi}$ above roughly 10 GeV, the cross-section is larger than current
bounds, see Figure \ref{Fig:Result}, and such chiral EW dark matter
is therefore ruled out as a viable dark matter candidate.
Note that for $m_{\chi}$ less than about 10 GeV (and down to about 2 eV, at which point the
dark matter ceases to be ``cold''), the direct detection cross-section
is not larger than the experimental bound.
However, since these particles couple to the $Z$-boson, the $Z$ could have decayed
into them.
The precise CERN LEP measurement of the invisible decay of the $Z$-boson rules out this
possibility.

\section{Non-Chiral Dark Matter} \label{Sec:Non-chiral Dark Matter}

Non-chiral, or vector, matter is different from chiral matter in
that an explicit mass term in the Lagrangian is allowed.
Even though, a priori, there is nothing that protects this explicit mass
term from being large, its size can nevertheless naturally be on the order
of the EW scale.
This may happen if, for example, the underlying high-scale theory has a global chiral
symmetry that is spontaneously broken at the EW scale, but that forbids an explicit mass
term at higher scales.

Non-chiral matter is not subject to the same tight constraints
from EW precision measurements as is chiral matter.
This is because there is no renormalizable coupling to the Higgs field.
Although there is a higher dimensional (non-renormalizable) coupling to
the Higgs, this does not cause any conflict with EW precision
measurements.
Instead, this coupling implies that non-chiral matter gains some small
fraction of its mass from EWSB.
It will be seen that the fraction of the dark matter particle's mass that
comes from EWSB is useful characterization of the dark matter's direct
detection cross-section.
This will be discussed further in \S \ref{Sec:singlet DM}.

Stability and electric neutrality are basic requirements of any dark
matter particle.
Since massive non-chiral representations are allowed to carry
conserved quantum numbers, which prohibits their mixing with
Standard Model fermions, the lightest state of such an additional
representation can indeed be stable.
Moreover, such representations contain both new neutral and new charged
particles.
The charged particles are several hundred MeV \emph{heavier} than
the neutral particles due to EWSB.
Intuitively one can understand the mass difference as arising from
different one-loop corrections to the masses and wave-functions: the
charged components receive corrections from both virtual photons and
Z-bosons in the loop, whereas the neutral components receive
corrections only from virtual Z-bosons \cite{Thomas:1998wy}.
This means that the lightest state of an additional massive non-chiral
representation can also be expected to be neutral.

It is useful to divide non-chiral representations up further into
real and complex representations.
Each of these will now be discussed by focusing on an explicit example.

\subsection{Real representations of $SU(2)_L \times U(1)_Y$}\label{Sec:Non-chiral Dark Matter:real
reps}

If the dark matter particle is part of a real representation of
$SU(2)_L$ $\times$ $U(1)_Y$, then its hypercharge, $Y$, must be
zero.
Since the charge, $Q$, of the dark matter must be zero, this also
implies $T_3 = Q - \frac{1}{2} Y = 0$.
The dark matter particle, now a Majorana fermion, therefore does
not couple to the $Z$-boson, and there is no coherent tree-level
scattering off nuclei.
This makes it ``safe'' from the current experimental bounds.

As an example, consider the dark matter to be part of an
$SU(2)_L$ triplet with zero hypercharge,
\begin{equation}
L = \left(\begin{array}{c} L^+ \\ L^0 \\ L^- \end{array}\right).
\end{equation}
Here the neutral component $L^0$ is a possible dark matter
candidate.  The explicit mass term in the Lagrangian is given by
\begin{equation}
{\cal L} \supset -\frac{m}{2} (2L^+L^- + L^0L^0).
\end{equation}

The non-renormalizable operator that, after EWSB, splits the mass of
the neutral components from the mass of the charged components by
several hundred MeV is given by \cite{Thomas:1998wy}
\begin{equation}
{\cal L} \supset \epsilon^{abc} L^a L^b H^{\dagger}T^c H,
\end{equation}
where the $T^a$, $a=1,2,3$, are the $SU(2)_L$ generators,
and $H$ is the Standard Model Higgs field.

The interactions of $L^0$ with the Standard Model gauge bosons and
the charged fields $L^{\pm}$ are given by
\begin{equation}\label{Eqn:L0_DM_couplings_to_other_particles}
g W_{\mu}^+ (-L^{+\dagger} \bar{\sigma}^{\mu} L^0 + L^{0\dagger}
\bar{\sigma}^{\mu} L^-) + g W_{\mu}^- (-L^{0\dagger} \bar{\sigma}^{\mu} L^+
+ L^{-\dagger} \bar{\sigma}^{\mu} L^0).
\end{equation}
Two-component spinor notation for the dark matter is employed
throughout this paper, while four-component Dirac notation will be
used below for the quark fields (in equation
(\ref{Eqn:L0_DM_couplings_to_other_particles}), $\sigma^{\mu} =
(I_2,\vec{\sigma})$ and $\bar{\sigma}^{\mu} = (I_2,-\vec{\sigma})$,
where $\vec{\sigma}$ are the usual Pauli matrices).

Note the absence of any coupling of the neutral component $L^0$ to
the $Z$-boson.
This means there is no tree-level scattering for $L^0$ off nuclei,
making this a viable dark matter candidate.
There is, however, an irreducible one-loop coupling to nucleons, which
will be discussed in \S \ref{Sec:Direct detection:non-chiral DM}.

The particle $L^0$ behaves like a wino-like lightest supersymmetric
particle (LSP) found in the Minimal Supersymmetric Standard Model
(MSSM).
Assuming that $L^0$ makes up all of the dark matter in the
universe, it may be shown that it must have a mass of about
\begin{equation}
m_{L^0} \simeq 2 ~\textrm{TeV}
\end{equation}
\noindent to give the correct dark matter relic density.
This mass was estimated from Figure 4 in \cite{Profumo:2004at}.
Non-perturbative electroweak corrections to the dark matter annihilation
cross-section as included in \cite{Cirelli:2007xd}
require the dark matter to have a mass of about 2.7 TeV to obtain
the correct relic density.

It is interesting to note that if $L^0$ makes up most of the
dark matter component in the universe, it will most likely be
very difficult to detect at the LHC.
Although a detailed collider study is beyond the scope of this paper,
the following comments are meant to give an indication of this difficulty.
Since the $L^{\pm,0}$ are heavy and weakly interacting, their production
cross-sections are small.  They may be very roughly estimated to be
on the order of $10^{-5}-10^{-4}$ pb, as may be extrapolated from Figure 2
in \cite{Cheung:2005ba}, which shows the production cross-section
for the related wino-like neutralinos and charginos in the MSSM.
Moreover, the charged states $L^{\pm}$ are split from the neutral
state $L^0$ only by a small amount, so that even though they produce
ionizing charged tracks, they do so only within the inner portion
of the detector, before they each decay into the neutral state
by emitting a soft pion \cite{Thomas:1998wy}.
The missing energy from the two neutral particles escaping the
detector balances, so that there is not much visible missing
energy.
At the LHC it is very difficult to trigger on this, and such
dark matter particles will thus be extremely difficult to detect
at the LHC.
It is possible but unlikely that a detailed collider study will
change this conclusion.

\subsection{Complex representation of $SU(2)_L \times U(1)_Y$}\label{Sec:Non-chiral Dark Matter:complex
reps}

If the dark matter particle is part of a complex representation of
$SU(2)_L$ $\times$ $U(1)_Y$, then its hypercharge is nonzero.
Since the charge of the dark matter must be zero, $T_3 = -\frac{1}{2} Y$.
The dark matter particle, now a Dirac fermion, therefore couples to the
$Z$-boson at tree-level.
In the notation of equation (\ref{Eqn:Goodman-Witten}), $Y_L = Y_R \equiv Y$,
and the cross-section per nucleon for scattering off nuclei is given
by
\beq \label{Eqn:Goodman-Witten complex reps} \sigma \simeq
\frac{G_F^2}{2\pi} m^2_{\chi
N}\frac{1}{A^2}\Big[(1-4\sin^2\theta_W)Z - (A-Z)\Big]^2 Y^2. \eeq
For the CDMS experiment, using Germanium, the scattering
cross-section per nucleon then becomes
\beq
\sigma \simeq 2 \times 10^{-39}~Y^2 \textrm{ cm}^2,
\eeq
which is experimentally ruled out.

If this tree-level coupling of the dark matter particle to the $Z$-boson
can be avoided or at least suppressed, this type of dark matter
again becomes viable.
This can be achieved for example by adding additional matter,
cf.~\cite{Hall:1997ah,TuckerSmith:2004jv,Schuster:2005ck,Belanger:2007dx,D'Eramo:2007ga}.
In the case of dark matter that is a doublet of $SU(2)_L$, however,
it can be achieved by a non-renormalizable operator that couples the
dark matter particle to the Higgs.

The example of two $SU(2)_L$ doublets of opposite hypercharge will now be
discussed in detail.
Denote the two $SU(2)_L$ doublets by
\begin{equation}
L_1 = \left( \begin{array}{c} L_1^0 \\ L_1^-
\end{array} \right) \hspace{1.5cm} L_2 = \left( \begin{array}{c} -L_2^+ \\
L_2^0\end{array} \right),
\end{equation}
where $L_1$ has hypercharge $Y=-1$, and $L_2$ has hypercharge $Y=+1$.
The explicit mass term in the Lagrangian is given by
\begin{equation}
{\cal L} \supset -m L_1 L_2,
\end{equation}
where the $SU(2)_L$ indices are contracted as $\epsilon_{\alpha\beta}L_1^{\alpha}L_2^{\beta}$.
The neutral components of each doublet together form a neutral Dirac fermion.

There is an accidental $U(1)_{L_1 L_2}$ symmetry under which $L_1$ and $L_2$
transform opposite to each other.
This symmetry requires the neutral components to be part of a
Dirac fermion, and thus allows the tree-level scattering off nuclei via
$Z$-boson exchange.
An operator which violates this symmetry can, however, split the Dirac state
into a pseudo-Dirac state, which consists of two Majorana fermions that
have a tiny mass splitting.
This splitting can substantially suppress the tree-level scattering.

The non-renormalizable operator that, after EWSB, splits the mass of the neutral
components from the mass of the charged components by several hundred MeV is given by
\begin{equation}
{\cal L} \supset L_2 T^a L_1 H^{\dagger} T^a H,
\end{equation}
where the $T^a$ are the $SU(2)$ generators \cite{Thomas:1998wy}.
This operator, however, only affects the splitting of the charged states
from the neutral states.
Since it does not violate the $U(1)_{L_1 L_2}$ symmetry, it does not
affect the neutral Dirac state, whose scattering off nuclei remains unchanged.

However, a non-renormalizable operator that does violate the $U(1)_{L_1 L_2}$ symmetry
is given by
\begin{equation}
{\cal L} \supset -\frac{c}{M}(L_1 H)(L_1 H) + h.c. -
\frac{c^*}{M}(L_2 H^c)(L_2 H^c) + h.c.,
\end{equation}
where brackets indicate that the $SU(2)_L$ indices are
contracted, $H^c = i\sigma_2 H^*$, and $H$ has been assigned hypercharge $Y=-1$.
The scale $M$ is some high mass scale at which this operator is generated, and
$c$ is an $O(1)$ coefficient.
Note that in writing down this term, the discrete symmetry
$L_1\leftrightarrow (L_2)^{^c}$ was assumed, so that the
coefficients are the same up to complex conjugation (removing this assumption
leaves unchanged the main conclusion, namely that the neutral Dirac state will
be split).
This operator only exists for dark matter that has hypercharge $|Y|=1$.

Once the Higgs field obtains a vacuum expectation value, $v$, and EW symmetry has been
broken, the neutral components get an additional contribution to the mass, which can be written as
$\delta =\frac{c}{M}v^2$.
$M$ will have to be large enough to ensure $|\delta| \ll m$.
Including corrections up to
$\mathcal{O}\Big(\frac{\textrm{\footnotesize{Im}}\delta}{m}\Big)$ or
$\mathcal{O}\Big(\frac{\textrm{\footnotesize{Re}}\delta}{m}\Big)$, the mass term may be
written as
\begin{eqnarray}
-\frac{1}{2}\left(\begin{array}{cc} L_1^0 & L_2^0
\end{array} \right) \left(\begin{array}{cc} \delta & m \\ m &
\delta^*
\end{array}\right) \left(\begin{array}{c} L_1^0 \\ L_2^0
\end{array}\right) \hspace{5.5cm} & & \nonumber \\
= -\frac{1}{2}\left(\begin{array}{cc} \chi_2 & \chi_1
\end{array} \right) \left(\begin{array}{cc} m+\textrm{\footnotesize{Re}}\delta& 0 \\ 0 &
m-\textrm{\footnotesize{Re}}\delta
\end{array}\right) \left(\begin{array}{c} \chi_2 \\ \chi_1
\end{array}\right),
\end{eqnarray}
where the neutral mass eigenstates are given by
\begin{eqnarray}
\chi_1 & \simeq &
\frac{i}{\sqrt{2}}\Bigg(\Bigg(-1+\frac{1}{2}\frac{\textrm{\footnotesize{Im}}\delta}{m}\Bigg)
L_1^0 +
\Bigg(1+\frac{1}{2}\frac{\textrm{\footnotesize{Im}}\delta}{m}\Bigg)
L_2^0
\Bigg) \\
\chi_2 & \simeq &
\frac{1}{\sqrt{2}}\Bigg(\Bigg(1+\frac{1}{2}\frac{\textrm{\footnotesize{Im}}\delta}{m}\Bigg)
L_1^0 +
\Bigg(1-\frac{1}{2}\frac{\textrm{\footnotesize{Im}}\delta}{m}\Bigg)
L_2^0 \Bigg)
\end{eqnarray}
These are the two Majorana fermions that make up the pseudo-Dirac state.
Ignoring higher order corrections, the mass eigenstates may also
be written as
\begin{eqnarray}\label{Eqn:mass eigenstates}
\chi_1 & \simeq & \frac{i}{\sqrt{2}} (-L_1^0 + L_2^0),
\hspace{1.5cm} m_1 = m -
\textrm{\footnotesize{Re}}{\delta} \\
\chi_2 & \simeq & \frac{1}{\sqrt{2}} (L_1^0 + L_2^0), \hspace{1.9cm}
m_2 = m + \textrm{\footnotesize{Re}}{\delta}.
\end{eqnarray}
Here $\chi_1$, the lighter of the two Majorana particles,
is the dark matter particle.
It behaves like a higgsino-like LSP found in the MSSM.
Assuming that $\chi_1$ makes up all of the dark matter in the universe,
it must have a mass of about
\begin{equation}
m_{\chi_1} \simeq 1~\textrm{TeV},
\end{equation}
to give the correct dark matter relic density.
This mass was estimated from Figure 4 in \cite{Profumo:2004at}.
Non-perturbative electroweak corrections are negligible as discussed in
\cite{Cirelli:2007xd}.

At lowest order, the couplings among the neutral fields, $\chi_1$ and $\chi_2$,
and the charged fields, $L_1^-$ and $L_2^+$, are given by
\begin{eqnarray}\label{Eqn:chi_1_DM_couplings_to_other_particles}
& &
L_1^\dagger(i\bar{\sigma}^{\mu}\partial_{\mu})L_1+L_2^\dagger(i\bar{\sigma}^{\mu}\partial_{\mu})L_2
+ g
W^+_{\mu}\Big[\frac{1}{2}(\chi_2^\dagger-i\chi_1^\dagger)\bar{\sigma}^{\mu}L_1^{-}
\nonumber \\ & &  - \frac{1}{2}{L_2^{+}}^\dagger
\bar{\sigma}^{\mu}(\chi_2-i\chi_1)\Big] + g
W^{-}_{\mu}\Big[-\frac{1}{2}(\chi_2^\dagger+i\chi_1^\dagger)\bar{\sigma}^{\mu}L_2^{+}
+ \frac{1}{2}{L_1^{-}}^\dagger \bar{\sigma}^{\mu}(\chi_2+i\chi_1)\Big]
\nonumber \\ & & + \frac{g}{\cos\theta_W} Z_{\mu}\Big[{L_1^{-}}^\dagger
\bar{\sigma}^{\mu}(-\frac{1}{2}+\sin^2\theta_W)L_1^{-} + {L_2^{+}}^\dagger
\bar{\sigma}^{\mu}(\frac{1}{2}-\sin^2\theta_W)L_2^{+} \nonumber \\
& & + \frac{i}{2}(\chi_2^\dagger\bar{\sigma}^{\mu}\chi_1 -
\chi_1^\dagger\bar{\sigma}^{\mu}\chi_2)\Big]+ e
A_{\mu}\Big[-{L_1^-}^\dagger\bar{\sigma}^{\mu} L_1^- +
{L_2^+}^\dagger\bar{\sigma}^{\mu} L_2^+\Big].
\end{eqnarray}
Including the next higher order correction, the coupling
of the dark matter to the $Z$-boson becomes
\begin{equation}\label{Eqn:higher_order_correction_Z-boson}
\frac{g}{2\cos\theta_W}
Z_{\mu}\Big[i(\chi_2^\dagger\bar{\sigma}^{\mu}\chi_1 -
\chi_1^\dagger\bar{\sigma}^{\mu}\chi_2) +
\frac{\textrm{\footnotesize{Im}}\delta}{m}
(\chi_2^\dagger\bar{\sigma}^{\mu}\chi_2 -
\chi_1^\dagger\bar{\sigma}^{\mu}\chi_1)\Big].
\end{equation}
Equations (\ref{Eqn:chi_1_DM_couplings_to_other_particles}) and
(\ref{Eqn:higher_order_correction_Z-boson}) show that the
only coupling $\chi_1$ has to itself at tree-level is suppressed by
a factor of $\frac{\textrm{\footnotesize{Im}}\delta}{m}$.
The dominant coupling of $\chi_1$ is to $\chi_2$, and it is
possible for $\chi_1$ to scatter \emph{inelastically} off nucleons
via $Z$-boson exchange ($\chi_1 \to \chi_2$).
This inelastic scattering will be kinematically inaccessible if the mass
splitting between $\chi_1$ and $\chi_2$ ($\sim$ 2\,Re($\delta$)) is
large enough.
Since the typical recoil energies of the nuclei in the
detector are expected to be on the order of a few 10's of keV, a splitting
of a few 10's of keV is required in order to forbid the inelastic scattering
via $Z$-boson exchange\footnote{The
question of whether the scattering is kinematically allowed or not
depends critically on the mass of the nucleus in the detector.
It is thus possible to carefully choose $\delta$ in such a way that
scattering will take place in a heavier target such as NaI used by
DAMA, but not in a lighter target such as Ge used by CDMS. The
possibility of using this to explain the DAMA signal, in the absence
of a signal by CDMS and others, was discussed in
\cite{Smith:2001hy,TuckerSmith:2004jv}. (The fact that the dark
matter in the halo would follow a Maxwell-Boltzmann distribution of
velocities complicates, but does not invalidate, the statements just
made.)} \cite{Jungman:1995df, Lewin:1995rx}.
This means that $\frac{\textrm{\footnotesize{Im}}\delta}{m}$ can be as
small as $\sim$ $10^{-7} - 10^{-8}$, so that the cross-section for the
scattering of $\chi_1$ to $\chi_1$ off nuclei is suppressed by a factor of
$\Big(\frac{\textrm{\footnotesize{Im}}\delta}{m}\Big)^2$ $\sim$
$10^{-14} - 10^{-16}$, which ensures it lies well below the current
experimental bound.
Note also that this requires the scale of the new physics which generates
the operator that breaks the $U(1)_{L_1L_2}$ symmetry to be roughly
$M\lesssim$ $10^8 - 10^9$ GeV.

For appropriate values of the mass splitting the dark matter can
therefore not scatter off the nuclei at tree-level.
This makes it ``safe'' from current experimental bounds.
There is, however, again an irreducible one-loop coupling to nucleons,
which will be discussed in \S \ref{Sec:Direct detection:non-chiral DM}.

It should be noted that $\chi_1$ will most likely be extremely
difficult to detect at the LHC.
The reasoning is similar to that mentioned at the end of
\S \ref{Sec:Non-chiral Dark Matter:real reps} for the case of
the $SU(2)_L$ triplet with zero hypercharge.
The LHC production cross-section of $\chi_1$ here is only marginally larger (since
it is less massive), about $10^{-4}-10^{-3}$ pb.  This was estimated from Figure 2
in \cite{Cheung:2005pv}, which shows the production cross-section for
the related higgsino-like neutralinos and charginos in the MSSM.
Moreover, the direct production of this type of dark matter and the associated charged
particles will again only give rise to signals that are very difficult to trigger on at
the LHC.
Their associated production with jets, for example, has a cross-section that is
too small to be visible above background events (see \cite{Chattopadhyay:2005mv},
which looked at collider signatures for a higgsino-like lightest supersymmetric
particle).
The non-chiral dark matter proposed in this paper thus seems to be
extremely difficult to detect at the LHC.
Although a detailed LHC collider study is beyond the scope of this paper,
it seems unlikely that it would change this conclusion.

\subsection{Direct detection of non-chiral dark matter}\label{Sec:Direct detection:non-chiral DM}

\begin{figure}\begin{center}\includegraphics[scale=0.8]{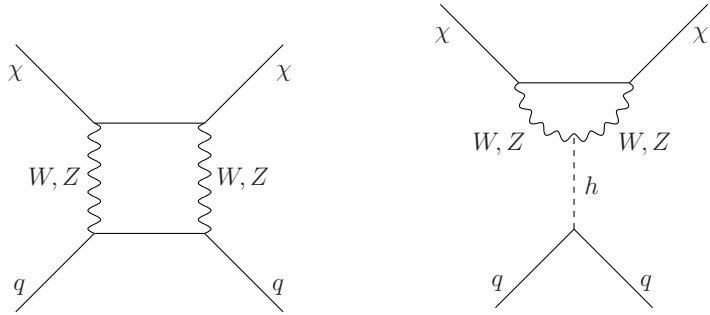}
\caption{\small{Feynman diagrams for the irreducible one-loop couplings
between non-chiral dark matter and quarks. For real representations
of $SU(2)_L$ $\times$ $U(1)_Y$, there are only $W$-bosons within the
loop. For complex representations of $SU(2)_L$ $\times$ $U(1)_Y$,
there are both $W$- and $Z$-bosons in the loop. The symbol $h$
denotes the Standard Model Higgs boson, $\chi$ denotes the dark
matter particle, and $q$ refers to quarks.
There is also a cross-diagram for the diagram on the left
which needs to be included.}
\label{Fig:Feynman1}}
\end{center}
\end{figure}

The previous two subsections considered non-chiral dark matter
that is either a real or a complex representation of
$SU(2)_L$ $\times$ $U(1)_Y$.
For real representations, there is no tree-level coupling between
the dark matter and the nuclei.
For complex representations, the tree-level coupling is completely
negligible, if the Dirac state has been appropriately split into
a pseudo-Dirac state.
Although the absence of any tree-level coupling allows non-chiral
dark matter particles to be consistent with current experimental
limits, there is an irreducible one-loop coupling which is large
enough for it to be detectable in future direct detection
experiments.\footnote{For indirect dark matter detection
rates and for prospects of detecting the associated charged particles
among the ultra-high energy cosmic rays see \cite{Cirelli:2007xd}.}
These irreducible one-loop couplings are given in Figure \ref{Fig:Feynman1}.

For real representations, the one-loop diagrams involve the
$W$-bosons, but not the $Z$-boson.
As an explicit example, consider the $SU(2)_L$ triplet with zero
hypercharge ($L^0$).
Its couplings to the $W$-bosons and to the additional charged states
($L^{\pm}$) are given in equation (\ref{Eqn:L0_DM_couplings_to_other_particles}).
The effective Lagrangian for the coherent interaction between the dark matter and the
quarks is
\begin{eqnarray}\label{Eqn:W-contribution_for_L0}
& & 4\, \alpha_2^2 \,\pi \,\sum_{q} \Bigg[\;\frac{1}{8}\, f_I^W
(m_W/m_{L^0}) \,\frac{1}{m_W m_h^2} \,\Big(L^0L^0 + L^{0\dagger}L^{0\dagger}\Big) \,m_q \bar{q} q \nonumber \\
& & \hspace{2cm}+\,\frac{1}{12}\,
f_{II}^W(m_W/m_{L^0}) \,\frac{1}{m_W^3 m_{L^0}} \,(L^0 iD^{\mu}
\sigma^{\nu}L^{0\dagger} + L^{0\dagger} iD^{\mu} \bar{\sigma}^{\nu}L^0) \,\times
\nonumber \\  & & \hspace{3cm} \bar{q}(\gamma_{\mu}iD_{\nu}+\gamma_{\nu}
iD_{\mu}-\frac{1}{2}g_{\mu\nu}i\Dslash)q \;\Bigg].
\end{eqnarray}
This result\footnote{The result for the one-loop computation agrees on-shell
with \cite{Cirelli:2005uq} for $m_W/m_{L^0}\to 0$, although here the operator
$\Big(\frac{1}{2}L^0L^0 + \frac{1}{2}L^{0\dagger}L^{0\dagger} \Big)\,\bar{q}i\Dslash q$
is found to vanish, and the coefficient of the twist-two operator is a factor of two
larger than in \cite{Cirelli:2005uq}.  The results of this paper do not agree off- or on-shell with
\cite{Hisano:2004pv}, who considered wino-like and higgsino-like lightest supersymmetric
particles in the MSSM.
Since the result agrees on-shell with \cite{Cirelli:2005uq}, the final cross-sections
calculated in this paper are also very similar in magnitude.
(It is more difficult to numerically compare the cross-sections with those of
\cite{Hisano:2004pv} since their's is dependent on various MSSM parameters.)}
was obtained by assuming that the momentum carried by
the quarks in the Feynman diagram on the left in Figure
\ref{Fig:Feynman1} is small but non-zero;  in the Feynman diagram on
the right the momentum of the quarks was set to zero, and therefore
no momentum was assumed to flow through the Higgs propagator.
The functions $f^W_{I}$ and $f^W_{II}$ are given by
\begin{eqnarray}
f_I^W (x) & = &
\frac{1}{3\pi}\Big(\frac{12-12x^2+2x^4}{\sqrt{4-x^2}}\arctan(\frac{1}{x}\sqrt{4-x^2})
\nonumber  \\ & & + 2x + (4x-x^3) \ln x^2\Big) \label{Eqn:f_W_I} \\
f_{II}^W (x) & = &
\frac{1}{4\pi}\Big(\frac{16+12x^2-12x^4+2x^6}{\sqrt{4-x^2}}\arctan(\frac{1}{x}\sqrt{4-x^2})
\nonumber \\ & & - 5x + 2x^3 + (4x^3-x^5) \ln x^2\Big).
\label{Eqn:f_W_III}
\end{eqnarray}
These functions have been normalized to equal one in the limit
$x\rightarrow 0$.
This is a useful normalization since here $x\equiv m_W/m_{L^0} \ll 1$.

For higher dimensional representations, there is an additional
factor in equation (\ref{Eqn:W-contribution_for_L0}).
For an $n$-tuplet of $SU(2)_L$ with zero hypercharge this additional
factor is given by $(n^2-1)/8$.

For complex representations, the one-loop diagrams involve the $W$-
and $Z$-bosons.
As an explicit example, consider the dark matter candidate from two
$SU(2)_L$ doublets of opposite hypercharge ($\chi_1$).
Its couplings to the $W$- and $Z$-bosons, to the
additional charged states $L_1^-$ and $L_2^+$, and to the slightly
heavier neutral state $\chi_2$ are given in equation
(\ref{Eqn:chi_1_DM_couplings_to_other_particles}).
The effective coherent interaction between the dark matter and the
quarks due to $W$-bosons in the loop is given by equation
(\ref{Eqn:W-contribution_for_L0}) by replacing $L^0$ with $\chi_1$
and by including a factor of 1/4 which multiplies the whole
equation.
The effective Lagrangian for the coherent interaction between the dark matter
and the quarks due to a $Z$-boson in the loop is given by\footnote{Comparing
the result found here with the on-shell result given in the published version of
\cite{Cirelli:2005uq} for $m_Z/m_{L^0}\to 0$ the following discrepancy is found: the
Higgs contribution is a factor of 3 smaller here, and (\emph{on-shell}) the factor of
$3(c_V^q)^2$ appearing in \cite{Cirelli:2005uq} for the box-diagram is here
found to be $(c_A^q)^2$.}
\begin{eqnarray}\label{Eqn:Z-contribution}
& & \frac{\alpha_2^2 \pi}{\cos^4\theta_W} \,\sum_{q}
\Bigg[\;\Bigg(-\frac{1}{16} \,f_I^Z (m_Z/m_{\chi_1}) \,\frac{{(c_V^q)}^2 -
{(c_A^q)}^2}{m_Z^3} \,+ \,\frac{1}{16}\, f_{II}^Z(m_Z/m_{\chi_1})\, \frac{1}{m_Z m_h^2}\Bigg)  \nonumber \\
& & \hspace{3cm}\times\Bigg(\chi_1\chi_1 + \chi_1^\dagger\chi_1^\dagger \Bigg)
m_q \bar{q} q \nonumber \\
& & \hspace{2.35cm} + \, \frac{1}{24}\,
f_{III}^Z(m_Z/m_{\chi_1}) \, \frac{{(c_V^q)}^2 + {(c_A^q)}^2}{m_Z^3
m_{\chi_1}}\, \Bigg(\chi_1 iD^{\mu}\sigma^{\nu}\chi_1^\dagger + \chi_1^\dagger iD^{\mu}
\bar{\sigma}^{\nu}\chi_1\Bigg) \nonumber \\
& & \hspace{3cm} \times \bar{q}\Bigg(\gamma_{\mu}iD_{\nu}+\gamma_{\nu}
iD_{\mu}-\frac{1}{2}g_{\mu\nu}i\Dslash\Bigg)q \;\Bigg]
\end{eqnarray}
where the functions $f^Z_{I}$, $f^Z_{II}$ and $f^Z_{III}$ are given by
\begin{eqnarray}\label{Eqn:Z-functions}
f_I^Z (x) & = &
\frac{1}{\pi}\bigg(\frac{4-2x^2+x^4}{\sqrt{4-x^2}}\arctan\Big(\frac{1}{x}\sqrt{4-x^2}\Big)
+ x -\frac{1}{2} x^3 \ln x^2\bigg) \label{Eqn:f_Z_I} \\ \nonumber
f_{II}^Z (x) & = &
\frac{1}{\pi}\bigg(\frac{4+4x^2-2x^4}{\sqrt{4-x^2}}\arctan\Big(\frac{1}{x}\sqrt{4-x^2}\Big) \\
& & - 2x + x^3 \ln x^2\bigg) \label{Eqn:f_Z_II} \\ \nonumber
f_{III}^Z (x) & = &
\frac{1}{8\pi}\bigg(\frac{32+16x^2-32x^4+8x^6}{\sqrt{4-x^2}}\arctan\Big(\frac{1}{x}\sqrt{4-x^2}\Big)
\\ & & - 4x + 8x^3 + (8x^3-4x^5) \ln x^2\bigg). \label{Eqn:f_Z_III}
\end{eqnarray}
These functions have also been normalized to equal one in the limit $x\rightarrow 0$.
This is again a useful normalization since here $x\equiv m_Z/m_{\chi_1} \ll 1$.
On the quark line the coupling of the $Z$ boson to the quarks is given
by $-\frac{g}{\cos\theta_W} \gamma^{\mu} \frac{1}{2} (c_V^q - c_A^q\gamma^5)$,
where $c_V^q = T^3_q - 2 \sin^2\theta_W Q_q$, $c_A^q = T^3_q$, $Q_q$ is the
quark charge, and $T^3_q = +\frac{1}{2}~(-\frac{1}{2})$ for up (down)-type
quarks.

For higher dimensional complex representations, there are additional factors
in the $W$-bosons contribution in equation
(\ref{Eqn:W-contribution_for_L0}) and in the $Z$-boson contribution
in equation (\ref{Eqn:Z-contribution}).
For an $n$-tuplet of $SU(2)$ with $n=Y+1$, there is an additional factor of
$(n^2-(1-Y)^2)/16$ multiplying equation (\ref{Eqn:W-contribution_for_L0}).
However, if $n>Y+1$, then there are more charged states that the
dark matter particle can couple to, and the additional factor
multiplying equation (\ref{Eqn:W-contribution_for_L0}) is given by
$(n^2-(1+Y^2))/8$.
For an $n$-tuplet of hypercharge $Y$, the factor that needs to
multiply equation (\ref{Eqn:Z-contribution}) is given by $Y^2$.

The effective coupling between dark matter and the quarks involves
several operators at a scale of order $m_Z$ (which is
the value of the dominant momentum in the loops of the diagrams in
Figure \ref{Fig:Feynman1}).
These operators are the scalar operator $m_q\bar{q} q$, the trace
operator $\bar{q}i\Dslash q$ (which was found to vanish, but there is nothing that
in principle forces it to vanish), and the traceless twist-two operator
$\frac{1}{2}\bar{q}(\gamma_{\mu}iD_{\nu}+\gamma_{\nu}
iD_{\mu}-\frac{1}{2} g_{\mu\nu} i\Dslash )q$.
The traceless twist-two operator and trace operator are part of the
quark energy momentum tensor given by $\bar{q}\gamma^{(\mu}iD^{\nu)}q$.

The nucleon matrix element of the scalar operator $m_q\bar{q}q$ for light
quarks is \cite{Jungman:1995df,Ellis:2000ds}
\begin{equation}\label{Eqn:scalar operator}
\langle N| m_q \bar{q}q |N\rangle = f_{T_q}^N m_N \bar{N} N,
\end{equation}
\noindent where on the right hand side of the equation $N$ denotes
a nucleon, and
\begin{eqnarray}
f^p_{T_u} \simeq 0.020\pm 0.004, ~~f^p_{T_d} \simeq 0.026\pm 0.005, ~~f^p_{T_s} \simeq
0.118 \pm 0.062 ~ \nonumber
\\
f^n_{T_u} \simeq 0.014 \pm 0.003, ~~f^n_{T_d} \simeq 0.036 \pm 0.008, ~~f^n_{T_s}\simeq
0.118 \pm 0.062.
\end{eqnarray}
The main contribution comes from the strange quark content of the nucleon, which
also has the largest uncertainty.
Heavy quarks, $Q$, also contribute to the mass of the nucleon.
This can be derived by making use of the anomaly relating the heavy quarks
to the gluons \cite{Jungman:1995df},
\begin{equation}\label{Eqn:scalar operator heavy quark}
\langle N| m_Q \bar{Q}Q |N\rangle = \langle N|-\frac{\alpha_s}{12
\pi} G^a_{\mu\nu} G^{a\mu\nu} |N\rangle= \frac{2}{27} f^N_{TG} m_N
\bar{N}N,
\end{equation}
where,
\begin{equation}\label{Eqn:scalar operator heavy quark evaluated}
\frac{2}{27}f^N_{TG} = \frac{2}{27} \Big(1-\sum_{u,d,s}
f^N_{T_q}\Big) \simeq 0.062.
\end{equation}

\begin{figure}\begin{center}\includegraphics[scale=0.8]{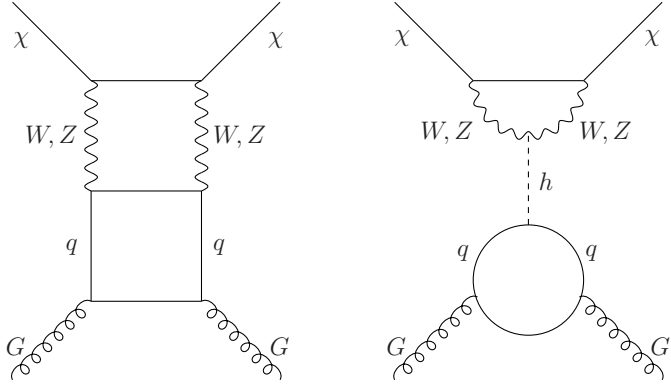}
\caption{\small{Feynman diagrams generating a coupling between dark
matter $\chi$ and gluons $G$.  For real representations of
$SU(2)_L$ $\times$ $U(1)_Y$, there are only $W$-bosons within the
loop. For complex representations of $SU(2)_L$ $\times$ $U(1)_Y$,
there are both $W$- and $Z$-bosons in the loop.  The symbol $h$
denotes the Standard Model Higgs boson and $q$ refers to quarks.}
\label{Fig:Feynman2}}
\end{center}
\end{figure}

Although the results in this paper suggest that the trace operator $\bar{q}i\Dslash q$ vanishes, and its
nucleon matrix element is therefore not needed, one may estimate it as follows:
The nucleon matrix element for light quarks may be estimated as
\begin{equation}\label{Eqn:trace operator light quarks}
\langle N| \bar{q}i\Dslash q |N\rangle = \langle N| m_q \bar{q}q|N\rangle.
\end{equation}
An accurate determination of the nucleon matrix element for the trace operator with heavy
quarks involves the calculation of higher loop diagrams as shown in
Figure \ref{Fig:Feynman2}, and is beyond the scope of this paper (see for example \cite{Drees:1992rr}).
Instead, as a crude approximation, equation (\ref{Eqn:trace operator light quarks}) may
also be used for the heavy quarks $Q$, together with (\ref{Eqn:scalar operator heavy quark}).

The twist-two quark operator is given by
\begin{equation}
{\cal O}^{(2)\mu\nu}_q =
\frac{1}{2}\bar{q}(\gamma^{\mu}iD^{\nu}+\gamma^{\nu}
iD^{\mu}-\frac{1}{2}g^{\mu\nu}i\Dslash)q.
\end{equation}
A linear combination of scale-dependent twist-two quark operators,
\beq\label{Eqn:lincombtwist2}
\sum_q \lambda_q [{\cal O}^{(2)}_q]_{m_{Z}^2},
\eeq
is generated at the scale $m_Z$, with coefficients $\lambda_q$ that may be read from
equations (\ref{Eqn:W-contribution_for_L0}) and (\ref{Eqn:Z-contribution}).
For the $W$-contribution to the scattering amplitude, the coefficients are the same
for all quarks $q$, but for the $Z$-contribution they differ for up- and down-type quarks.
Under QCD rescaling, the twist-two quark operator mixes with the twist-two gluon operator.
This may be taken into account by rewriting equation (\ref{Eqn:lincombtwist2}) as a linear
combination of operators that rescale multiplicatively \cite{Peskin:1995Book}.
One of these operators is the QCD energy-momentum tensor $T^{\mu\nu}$
\begin{equation}\label{Eqn:energymomentumtensor}
T^{\mu\nu} = \sum_q {\cal O}^{(2)\mu\nu}_q + {\cal O}^{(2)\mu\nu}_G,
\end{equation}
where ${\cal O}^{(2)}_G$ is the twist-2 gluon operator given by
\begin{equation}
{\cal O}^{(2)\mu\nu}_G =
G^{a\rho\mu}G^{a\nu}_{\rho}-\frac{1}{4}g^{\mu\nu}G^{a\rho\sigma}G^{a}_{\rho\sigma}.
\end{equation}
Another operator that may be rescaled multiplicatively is
\beq\label{Eqn:orthogonaloperator}
{\cal O}^{\mu\nu}_{-} = \frac{16}{3}\sum_q {\cal O}^{(2)\mu\nu}_q - n_f \,{\cal O}^{(2)\mu\nu}_G,
\eeq
where $n_f$ is the number of active quark flavors ($n_f = 5$ at the scale $m_Z$).
In the case of the $W$-contribution, for which all $\lambda_q$ are the same,
equation (\ref{Eqn:lincombtwist2}) can be rewritten in terms of the operators
(\ref{Eqn:energymomentumtensor}) and (\ref{Eqn:orthogonaloperator}).
For the $Z$-contribution, however, $\lambda_q$ differs for up- and down-type
quarks, so that other operators that rescale multiplicatively are required.
These are flavor non-singlet combinations of the individual quark
operators ${\cal O}^{(2)}_{q_i} - {\cal O}^{(2)}_{q_j}$ that do not mix
with the gluon operator since the gluon contributions cancel out.

The linear combination of twist-two quark operators (\ref{Eqn:lincombtwist2})
can thus be rewritten in terms of operators whose QCD rescaling is simple.
The operators may then be rescaled down to low scales, so that
(\ref{Eqn:lincombtwist2}) may be written in terms of operators that are
evaluated at low scales.
The energy-momentum tensor $T^{\mu\nu}$ has zero anomalous dimension, whereas
${\cal O}^{\mu\nu}_{-}$ and the flavor non-singlet combinations
${\cal O}^{(2)}_{q_i} - {\cal O}^{(2)}_{q_j}$ have positive anomalous dimension
given by $\frac{\alpha_s}{3\pi}\big(\frac{16}{3}+n_f\big)$ and $\frac{16\alpha_s}{9\pi}$,
respectively.
This means that running to the infrared, $T^{\mu\nu}$ does not get renormalized whereas
the other operators, ${\cal O}^{\mu\nu}_{-}$ and ${\cal O}^{(2)}_{q_i} - {\cal O}^{(2)}_{q_j}$,
both \emph{decrease}.
The dominant contribution at low scales to the linear combination of twist-two
quark operators generated at $m_Z$ is thus from the quark
energy momentum tensor, whose contribution is known exactly.
The other contributions are subdominant, and may be estimated from the
parton distribution functions (PDFs); helpful for this is \cite{PDFs}.
The expression for (\ref{Eqn:lincombtwist2}), written in terms of the
operators evaluated at a lower scale, will not be reproduced here.
However, it was checked that for a lower scale equal to 1 GeV,
the subdominant contributions that require knowledge of the PDFs amount
to only about 17$\%$ in the case of the $W$-contribution and 14$\%$ in the
case of the $Z$-contribution (care was taken to decrease the active number
of quark flavors from five to four at the scale of the bottom quark mass
and from four to three at the scale of the charm quark mass).
This shows that the nucleon matrix element of the twist-two quark operator
can be estimated reliably.

The nucleon matrix element of the twist-two quark operators may be evaluated
by using the expression \cite{Jungman:1995df}
\begin{eqnarray}\label{Eqn:Twist-2-matrix-element}
\langle N(p)| {\cal O}^{(2)\mu\nu}_q |N(p)\rangle & = &
\frac{1}{m_N}\Big(p^{\mu}p^{\nu}-\frac{1}{4}m_N^2 g^{\mu\nu}\Big)
\nonumber \\
& & ~~~~\times \int_0^1 dx \,x\, \big(q(x,\mu^2)+\bar{q}(x,\mu^2)\big),
\end{eqnarray}
where $p^{\mu}$ denotes the momentum of the nucleon, and
zero momentum transfer was assumed.
The PDF $q(x,\mu^2)$ (or $\bar{q}(x,\mu^2)$) gives the probability density
of finding the quark $q$ (or anti-quark $\bar{q}$) in the nucleon with momentum
fraction $x$.
The integral denotes the second moment of the PDF, and one may define
\begin{equation}
q(2,\mu^2) = \int_0^1 dx \, x \, q(x,\mu^2).
\end{equation}
The PDFs depends on the scale $\mu$ at which the twist-two operator was generated,
so that here $\mu=m_Z$.
Using the website \cite{PDFs} and the results from the CTEQ group (CTEQ6M)
\cite{Pumplin:2002vw}, the second moment of the PDFs may be determined directly at this
scale (equivalently, $q(2,\mu^2)$ may be determined at $\mu=1$ GeV if the linear combination
of twist-two quark operators is first rescaled down to 1 GeV).
The second moment of the PDFs for the proton for $\mu = m_Z$ are given by
\begin{eqnarray}\label{Eqn:pdf's}
& & u(2) \simeq 0.221, ~~ \bar{u}(2) \simeq 0.034, \nonumber \\
& & d(2) \simeq 0.115, ~~ \bar{d}(2) \simeq 0.039, \nonumber \\
& & s(2) \simeq 0.026, ~~ \bar{s}(2) \simeq 0.026, \\
& & c(2) \simeq 0.019, ~~ \bar{c}(2) \simeq 0.019, \nonumber \\
& & b(2) \simeq 0.012, ~~ \bar{b}(2) \simeq 0.012, \nonumber \\
& & G(2) \simeq 0.47.\nonumber
\end{eqnarray}
$G(2)$ is the PDF of the gluon, which is not needed here.
For the neutron, the values of $u(2)$ and $\bar{u}(2)$ are
interchanged with $d(2)$ and $\bar{d}(2)$, respectively.

The nucleon matrix elements discussed above may now be used to write the
spin-independent effective Lagrangian for non-chiral dark
matter scattering off nucleons as
\beq\label{Eqn:effective Lagr nucleon scattering}
\mathfrak{L}_{\textrm{\scriptsize{eff, N}}}^{\chi} \simeq \mathcal{C} \, m_N
\, \Big(\frac{1}{2} \chi\chi + \frac{1}{2} \chi^\dagger \chi^\dagger\Big) \bar{N}N,
\eeq
where $\mathcal{C}$ is determined from equations (\ref{Eqn:W-contribution_for_L0})
and (\ref{Eqn:Z-contribution}) and using the nucleon matrix elements.
The cross-section for the non-chiral dark matter particle to scatter
off nuclei (normalized to a single nucleon) is then
\begin{equation}
\sigma_{N}^{\chi} = \frac{1}{\pi}\, \mu^2_{\chi N} \,m_N^2 \,\mathcal{C}^2,
\end{equation}
where $\mu^2_{\chi N}$ is the reduced mass of the nucleon and the dark matter.
The cross-section for a dark matter particle from an $SU(2)_L$ triplet
with $Y=0$ is roughly the same when scattering off a proton or a neutron,
and the average is given by
\begin{equation}\label{Eqn:cross-section triplet}
\sigma_N^{L^0} \simeq 1.9 \times 10^{-45}\, {\textrm{cm}}^{2}.
\end{equation}
The cross-section for a dark matter particle from two $SU(2)_L$ doublets
with opposite hypercharge $Y=\pm 1$, after splitting the Dirac state into
a pseudo-Dirac state, is also roughly the same when scattering
off a proton or a neutron, and the average is given by
\beq\label{Eqn:cross-section doublets}
\sigma_N^{\chi_1} \simeq 2.1 \times 10^{-46}\, {\textrm{cm}}^{2}.
\eeq
A Higgs mass of $m_h = 120$ GeV was assumed.
For higher dimensional representations there are additional factors which
increase the cross-section, as discussed below equations
(\ref{Eqn:W-contribution_for_L0}) and (\ref{Eqn:Z-contribution}).
For example, a quintuplet of $SU(2)_L$ with $Y=0$ has a cross-section
that is larger by a factor of 9 than the triplet cross-section, i.e.
$\sigma \simeq 3.9 \times 10^{-44}$ cm$^2$.

\begin{figure}\begin{center}\includegraphics[scale=0.71]{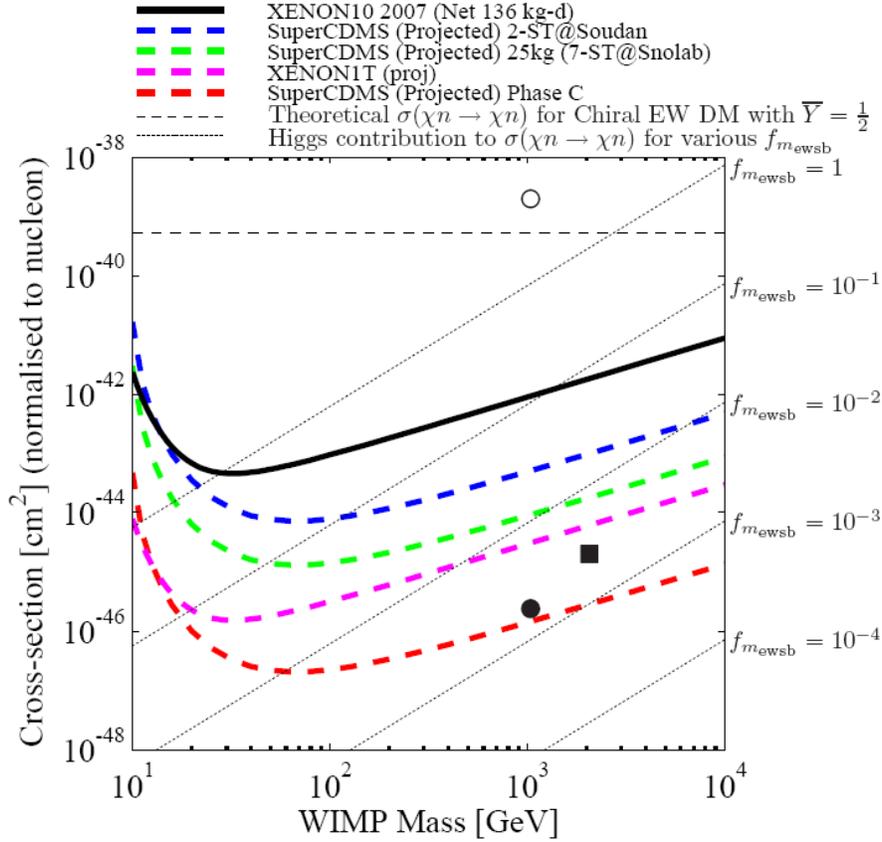}
\caption{\label{Fig:Result}\small{A comparison of the results presented
in this paper with current and projected experimental bounds for the cross-section of
dark matter scattering off a nucleon.
Shown are the current experimental upper bounds from XENON10 (solid black line)
\cite{Angle:2007uj}, and the projected upper bounds for SuperCDMS 2-ST at Soudan
(blue dashed line), SuperCDMS 25kg / 7-ST at Snolab (green dashed line),
XENON1T (magenta dashed line) and SuperCDMS Phase C (red dashed line)
\cite{Aprile:2005mz,Brink:2005ej,Schnee:2005pj}.
The dashed black horizontal line is the theoretical lower bound on the cross-section
for chiral electroweak dark matter scattering coherently off nuclei via the exchange
of a $Z$-boson, see \S\ref{Sec:Chiral DM}.
The black dot ({\large{$\bullet$}}) is the predicted cross-section for
a 1 TeV non-chiral dark-matter particle part of two $SU(2)_L$
doublets with opposite hypercharge (a complex representation of
$SU(2)_L$), assuming its coupling to the $Z$-boson is forbidden by splitting
the Dirac state into a pseudo-Dirac state; see
\S\ref{Sec:Non-chiral Dark Matter:complex reps}.
Without the latter assumption, the cross-section is given by the open circle
({\large{$\circ$}}) and would be ruled out.
The black square ({\scriptsize{$\blacksquare$}}) is the predicted cross-section
for a 2 TeV non-chiral dark matter particle part of an $SU(2)_L$
triplet with zero hypercharge (a real representation of $SU(2)_L$),
see \S\ref{Sec:Non-chiral Dark Matter:real reps}.
Dark matter from higher order real or complex representations has a larger
direct-detection cross-section than those represented by the black square or by the
black dot, respectively, see \S\ref{Sec:Non-chiral Dark Matter}.
The dotted diagonal lines represent the Higgs contribution to dark
matter scattering off nucleons for a range of magnitudes of the Higgs to dark matter
coupling.
This coupling also determines what fraction,
$f_{m_{\textrm{\tiny{ewsb}}}} \equiv m_{\textrm{\tiny{ewsb}}}/m_{\chi}$,
of the dark matter mass comes from electroweak symmetry breaking, and the
lines shown are for various $f_{m_{\textrm{\tiny{ewsb}}}}$.
The experimental results shown in this figure were obtained through
\cite{CDMS_graph_plotter}.}}
\end{center}
\end{figure}

Figure \ref{Fig:Result} shows the results for the cross-section and
how they compare to current experimental exclusion bounds,
as well as projected future bounds.
The current upper bound on the direct detection cross-section is roughly
two to three orders of magnitude higher than the calculated cross-sections in
(\ref{Eqn:cross-section triplet}) and (\ref{Eqn:cross-section doublets}),
respectively.
Interestingly, XENON1T will get close to the required sensitivity to see
an $SU(2)_L$ triplet with zero hypercharge and should be able to detect an
$SU(2)_L$ quintuplet with zero hypercharge, while SuperCDMS 25kg / 7-ST at
Snolab may not quite be able to detect the triplet, but will get close to
detecting the quintuplet.
Experiments planned for well into the future, such as the proposed SuperCDMS
``Phase C'' \cite{Brink:2005ej,Schnee:2005pj}, should be able to also probe the
required parameter space for the case of the two $SU(2)_L$ doublets with opposite
hypercharge.

\section{Higgs contribution to the Direct Detection \\ Cross-Section and Singlet Dark Matter}\label{Sec:singlet DM}

In this section, singlet dark matter will be discussed, and a useful
characterization of its direct detection cross-section will be given.
Dark matter that is a singlet under $SU(2)_L\times U(1)_Y$ does not have any
irreducible couplings to quarks, unlike the non-chiral dark matter discussed in
\S \ref{Sec:Non-chiral Dark Matter}.
It will be assumed that the singlet dark matter does not couple to the Higgs at the
renormalizable level and does not obtain a mass spontaneously.
Rather, the singlet will be allowed in the Lagrangian to have an explicit mass
term which is not associated with the EW scale.
Although there is no renormalizable coupling between the singlet and the
SM, no symmetries forbid the existence of a non-renormalizable interaction
generated by new physics beyond the SM at some high scale.
The gauge invariant operator coupling the dark matter
$\chi$ to the Higgs is an infinite sum of higher-dimensional operators,
\beq\label{Eqn:Hchichi}
\mathcal{L}_{h\chi\chi} = \frac{c_1}{\Lambda_1}\chi\chi H^{\dagger}H +
\frac{c_2}{\Lambda_2^3}\chi\chi (H^{\dagger}H)^2 + \ldots +
\frac{c_n}{\Lambda_n^{2n-1}} \chi\chi (H^{\dagger}H)^n + \ldots,
\eeq
where one Higgs field is replaced by the physical Higgs boson $h/\sqrt{2}$,
and all others acquire a vacuum expectation value of $v/\sqrt{2} \simeq 174$ GeV.
The $c_n$ are dimensionless coefficients and the $\Lambda_n$ are the scales
at which the higher dimensional operators are generated by new physics.

The Higgs-dark-matter coupling (\ref{Eqn:Hchichi}) is allowed more generally for
any non-chiral dark matter, whether it is a singlet or forms a non-trivial representation
of the EW gauge group.
For singlet dark matter, the coupling (\ref{Eqn:Hchichi}) is generated at a scale
$\Lambda_1$ by new physics.
For non-chiral dark matter with non-trivial EW quantum numbers, the coupling
is already generated at the EW scale by integrating out the $W$-bosons (and,
for complex representations, also the $Z$-boson), as shown in Figure
\ref{Fig:Feynman1} in \S \ref{Sec:Non-chiral Dark Matter}.

The existence of this Higgs to dark matter coupling also implies
the existence of additional contributions to the dark matter mass when all of
the Higgs fields in (\ref{Eqn:Hchichi}) acquire a vacuum expectation value.
This means that non-chiral dark matter obtains at least some of its mass from EWSB.
Denoting the dark matter mass by $m_{\chi}$ and the mass that is not associated with
EWSB by $m_0$, gives the relation
\beq\label{Eqn:Mass contribution}
m_{\chi} = m_0 + m_{\textrm{\footnotesize{ewsb}}},
\eeq
where $m_{\textrm{\footnotesize{ewsb}}} \simeq \frac{v^2}{2\Lambda_1} + \ldots$ is the
mass gained from EWSB.

The mass obtained by the dark matter from EWSB is a useful characterization
of the Higgs contribution to the direct detection cross-section.
The latter is given by
(see also \cite{McDonald:1993ex}, \cite{Kim:2006af})
\beq\label{Eqn:Singlet direct detection}
\sigma^{\chi}_N \simeq \frac{g^2}{4\pi m_W^2 m_h^4}\, \mu^2_{\chi N} \,m_N^2
    \,\Big(\sum_q f_{T_q}^N\Big)^2 \, g^2_{h\chi\chi},
\eeq
where $g_{h\chi\chi}$ $\simeq$ $c_1 v/2\Lambda_1$ $\simeq$
$m_{\textrm{\footnotesize{ewsb}}}/v$ is the Higgs to dark matter coupling,
and $f_{T_q}^N$ may be taken from equations
(\ref{Eqn:scalar operator}) $-$ (\ref{Eqn:scalar operator heavy quark evaluated}).
Evaluating the cross-section for $m_h\simeq$ 120 GeV gives
\beq\label{Eqn:Singlet direct detection evaluated}
\sigma^{\chi}_N \simeq 8\times 10^{-47} \, \mu^2_{\chi N} \,
    m_{\textrm{\tiny{ewsb}}}^2,
\eeq
or,
\beq\label{Eqn:Singlet direct detection evaluated 2}
\sigma^{\chi}_N \simeq 8\times 10^{-47} \, \mu^2_{\chi N} \, m_{\chi}^2 \,
    f_{m_{\textrm{\tiny{ewsb}}}}^2,
\eeq
where
\beq
f_{m_{\textrm{\tiny{ewsb}}}} \equiv \frac{m_{\textrm{\tiny{ewsb}}}}{m_{\chi}}
\eeq
is the dark matter mass fraction obtained from EWSB.
The cross-section is seen to be directly proportional to the square of this fraction.

The various dotted lines in Figure \ref{Fig:Result} show the cross-section for
$f_{m_{\textrm{\tiny{ewsb}}}}$ = 1, $10^{-1}$, $10^{-2}$ $10^{-3}$, and
$10^{-4}$, as well as the current experimental bounds. (Constraints on $m_{\chi}$
and $f_{m_{\textrm{\tiny{ewsb}}}}$ from the known dark matter relic density are
not included in the present discussion, but see for example
\cite{McDonald:1993ex}, \cite{Kim:2006af}).
These lines represent the Higgs contribution to the direct detection cross-section.
Modulo destructive interference with other contributions, they represent
the lower bounds of the direct detection cross-section also for non-chiral
dark matter that is not an EW singlet.\footnote{Note that for the $SU(2)_L$
triplet with zero hypercharge (\S \ref{Sec:Non-chiral Dark Matter:real reps}),
the Higgs contribution to the direct detection cross-section is about $8.7\times 10^{-47}$cm$^2$, so that
$f_{m_{\textrm{\tiny{ewsb}}}} \simeq 5.5\times 10^{-4}$ and $m_{\textrm{\tiny{ewsb}}}$ $\simeq$ 1.1 GeV.
For the $SU(2)_L$ doublets with opposite hypercharge (\S \ref{Sec:Non-chiral Dark Matter:complex reps}),
the Higgs contribution amounts to about $1.6\times 10^{-47}$cm$^2$, so that
$f_{m_{\textrm{\tiny{ewsb}}}} \simeq 4.5\times 10^{-4}$ and $m_{\textrm{\tiny{ewsb}}}$ $\simeq$ 0.5 GeV.}

If the dark matter is associated with new physics at the EW scale, the fraction
$f_{m_{\textrm{\tiny{ewsb}}}}$ should not be too small.
The current bound has ruled out dark matter with a mass heavier than
about 1 TeV and that obtains more than 10$\%$ of its mass from
EWSB.
SuperCDMS ``Phase C'' would be able to rule out dark matter with a
mass heavier than about 1 TeV and that obtains more than about $0.1\%$ of
its mass from EWSB.
This means that, assuming $c_1\sim\mathcal{O}$(1), SuperCDMS ``Phase C'' would
probe a scale of $\Lambda_1\sim\mathcal{O}$(30 TeV).
As the direct detection experiments probe ever smaller values of
$f_{m_{\textrm{\tiny{ewsb}}}}$, the absence of any direct detection signal
would make relevant the question of whether one should abandon the idea that
dark matter is associated with new physics at the EW scale.

\section{Conclusions} \label{Sec:Conclusion}

Fermion dark matter transforming under the electroweak gauge group
$SU(2)_L$ $\times$ $U(1)_Y$ was added to the standard model, and
the observational consequences at a direct detection experiment were
discussed.
Figure \ref{Fig:Result} summarizes the results.

Chiral electroweak dark matter is well known to be not a viable dark matter
candidate, as it has a spin-independent coupling to nuclei via the
$Z$-boson, which gives a cross-section that is ruled out by two to three
orders of magnitude.

Non-chiral dark matter from real representations of $SU(2)_L\times U(1)_Y$
has an irreducible one-loop spin-independent coupling to nuclei.
The triplet has a mass of about 2 TeV and a cross-section that is about two
order of magnitude below current experimental bounds.
A future experiment with a very large sensitivity, such as the proposed
XENON1T, is required to probe the relevant region of parameter space.
Higher order representations have a larger cross-section which makes it
easier to detect them.

Non-chiral dark matter from complex representation of $SU(2)_L\times U(1)_Y$
have a tree-level coupling to nuclei via $Z$-boson exchange, which would rule
it out unless this tree-level coupling can be suppressed somehow.
For two $SU(2)_L$ doublets with opposite hypercharge the tree-level coupling
can be suppressed by a dimension five operator that couples the Higgs to the
dark matter particle and is able to split the neutral Dirac state into a
pseudo-Dirac state.
The remaining irreducible one-loop coupling allows such a dark matter particle
to be detected at a very sensitive future planned direct detection experiment
such as SuperCDMS ``Phase C''.
Its mass is required to be about 1 TeV to reproduce the observed dark
matter relic density.

Although a detailed LHC collider study was not done in this paper,
non-chiral dark matter particles are most likely extremely difficult
to detect at the LHC.
The reason is that not many of them will be produced since they are not only
required to be heavy to reproduce the observed relic density, but they are
also weakly interacting.
This is in addition to the fact that they would not even provide a signal that
can easily be triggered on.

Non-chiral dark matter has a coherent coupling to the standard model fermions
through the Higgs field.
The existence of this coupling to the Higgs also means that at least some of
its mass is obtained from electroweak symmetry breaking.
Non-chiral dark matter from non-trivial representations of the
electroweak gauge group does indeed gain a small fraction, about $10^{-3}$,
of its mass from electroweak symmetry breaking.
For dark matter that is a singlet under the electroweak gauge group,
a non-renormalizable coupling to the Higgs could allow it to be detected at
a direct detection experiment (the singlet's dominant coupling to the Higgs
was assumed to be through a dimension five operator).
A useful characterization of the direct detection cross-section is
given by the fraction of mass that the dark matter particle obtains through
electroweak symmetry breaking, the amplitude being directly proportional to
this fraction.
The current experimental bound has ruled out dark matter with a mass
heavier than about 1 TeV and that obtains more than 10$\%$ of its mass from
EWSB.
SuperCDMS ``Phase C'' would be able to rule out dark matter with a
mass heavier than about 1 TeV and that obtains more than $0.1\%$ of its mass
from EWSB.
As the direct detection experiments probe ever more of the available parameter
space, the absence of any direct detection signal would at some point make
relevant the question of whether one should abandon the idea that dark matter
is associated with new physics at the EW scale.

\vspace{0.7cm}

\noindent {{\bf\Large Acknowledgements:}}

\vspace{0.2cm}

The author would like to thank S.~Thomas for suggesting this problem and
for many helpful and enlightening discussions.
The author would also like to thank M.~Cirelli, H.~Dreiner, J.F.~Fortin, N.~Sehgal, 
A.~Strumia, G.~Torroba, and K.~van den Broek for useful discussions or correspondence.
This research is supported by the Department of Physics and Astronomy
at Rutgers University.

\bibliographystyle{utcaps}
\bibliography{DM2.bbl}
%\bibliography{Bibliography}

\end{document}